\documentclass[twocolumn,prd,aps,superscriptaddress]{revtex4}

\usepackage{graphicx}
\usepackage{dcolumn}
\usepackage{bm}

\begin{document}
\voffset = 0.3 true in

\title{Hybrid Meson Potentials and the Gluonic van der Waals Force}

\author{Olga Lakhina}
\affiliation{
Department of Physics and Astronomy, University of Pittsburgh,
Pittsburgh PA 15260}

\author{Eric S. Swanson}
\affiliation{
Department of Physics and Astronomy, University of Pittsburgh,
Pittsburgh PA 15260}
\affiliation{
Jefferson Lab, 12000 Jefferson Ave,
Newport News, VA 23606}

\begin{abstract}
\vskip .3 truecm
The chromoelectric polarizability of mesons governs the strength of the gluonic
van der Waals force and therefore of non-quark-exchange processes
in hadronic physics. We compute the polarizability of heavy mesons with
the aid of lattice gauge theory and the Born--Oppenheimer adiabatic expansion. 
We find that the operator product expansion breaks down at surprisingly
large quarks masses due to nonperturbative gluodynamics and that previous conclusions concerning 
$J/\psi$--nuclear matter interactions and $J/\psi$ dissociation in
the quark-gluon plasma must be substantially modified.
\end{abstract}

\maketitle

\section{Introduction}

Although hadronic interactions are a central phenomenon of nuclear
and hadronic physics, very little can be said about them from first
principles and a microscopic description remains elusive.
One of the few attempts at describing hadronic interactions from QCD originated
more than twenty years ago with the operator product expansion (OPE) approach of
Peskin\cite{Peskin}. Peksin argued that the coupling of soft external gluons
to small (heavy quark) hadrons may be considered a short distance phenomenon
and therefore amenable to Wilson's operator product formalism. 
More recently, Luke, Manohar, and Savage \cite{LMS} have placed Peskin's argument in the
context of effective Lagrangians. Briefly,
in the absence of flavour exchange processes, the
interactions of hadronic matter are dominated by multigluon exchange
processes which may be described with an effective Lagrangian
at the compositeness scale,
$\Lambda_Q \sim r_Q^{-1} \sim \alpha_s(\Lambda_Q) m_Q$:

\begin{equation}
L^{(1)}_{eff} = \sum_v {1 \over \Lambda_Q^3}(P(v)^\dagger P(v) - V_\mu(v)^\dagger V(v)^\mu)
(c_E {\cal O_E} + c_B{\cal O}_B).
\label{L0}
\end{equation}

\noindent
Here the gluonic operators are ${\cal O}_E = -G^{\mu\alpha}G^\beta_\alpha v_\mu v_\beta$
and
${\cal O}_B = {1\over 2}G^{\alpha\beta}G_{\alpha\beta} -
G^{\mu\alpha}G^\beta_\alpha v_\mu v_\beta$;
$P(v)$ and $V_\mu(v)$ create pseudoscalar or vector mesons with four velocity $v^\mu$.
In the meson rest frame these operators reduce to $E^2$ and $B^2$ respectively.
The multipole formalism becomes exact in the large quark mass/small hadron
limit and 
the coupling constants (or Wilson coefficients) $c_E$ and $c_B$ may be interpreted as the chromoelectric and magnetic
polarizabilities of the heavy meson.

Luke, Manohar, and Savage used this formalism to estimate the binding energy
of quarkonium ($\Upsilon$ and $\Psi$) with nuclear matter\cite{LMS}. The gluonic
matrix elements were estimated with the aid of the scale anomaly and the
experimentally determined gluonic momentum fraction of the nucleon at the scale $\Lambda_Q$.
The final ingredient was Peskin's original estimate of the chromoelectric polarizability 

\begin{equation}
c_E = {14 \pi\over 3 (N_c^2-1)}
\label{ce}
\end{equation}

\noindent
(in the large $N_c$ limit). We note that theoretical uncertainty is introduced through the
choice of the compositeness scale, the strong coupling $\alpha_s$, and
the size of the meson. The final estimates of the binding energies were 
roughly 3 MeV for the $\Upsilon$ and 10 MeV for the $J/\psi$.
Subsequently, Brodsky and Miller\cite{BM} used this result to obtain 
a $J/\psi$--nuclear matter scattering length of $a_B = -0.24$ fm and a cross section of
roughly 7 mb at threshold. Brodsky and Miller also argued that multiple 
gluon exchange dominates the $J/\Psi$--nuclear matter interaction.

Finally, Kharzeev and Satz\cite{KS} have applied Peskin's results to the interaction of $J/\Psi$
with comoving matter in heavy ion collisions. They argue that the cross section is small
near threshold and that therefore collision-induced dissociation should not confound
the use of $J/\Psi$ suppression as a diagnostic for the formation of the quark-gluon 
plasma.

The chromoelectric polarizability has appeared in at least one other context.
Leutwyler has argued that nonperturbative level shifts in the heavy quarkonium spectrum
may be related to the product of the vacuum expectation value of the electric field 
pair density and the electric polarizability\cite{HL} (similar arguments have been made
with QCD sum rules\cite{MV}). In particular he states that the small size
of the heavy meson implies that quarks interact with slowly varying random chromofields.
The energy shift is then given by the expectation of the operator

\begin{equation}
\delta H = -P {\bf E} \cdot {\bf r} {1\over H_a-E_\phi} {\bf E}\cdot{\bf r} P
\label{leut}
\end{equation}
where $P$ projects onto mesonic states which are orthogonal to the meson, $E_\phi$ is the
mass of the heavy meson, ${\bf E}$ is the
chromoelectric field, and $H_a$ is the Hamiltonian which describes the interactions of
quarks in the colour octet state, 

\begin{equation}
H_a = 2m_Q + {p^2 \over m_Q} + {\alpha_s\over 2 N_c r}.
\end{equation}
The potential in $H_a$ is the perturbative expression for the interaction of a quark and an
anti-quark in the colour adjoint representation. The expectation value of $\delta H$ is
proportional to $c_E \langle E^2 \rangle$  and thus the electric polarizability
gives the strength of nonperturbative mass shifts (or, equivalently, the strength
of the nonlocal nonperturbative potential due to interactions with the gluon condensate).

It is clear that the value of the electric polarizability is crucial to all these 
conclusions. In the following, we carefully examine Peskin's computation of $c_E$
and conclude that its true value is roughly a factor of ten smaller than claimed.
More importantly, it will be shown that, in this application, {\it the 
operator product expansion is
never reliable in Nature} due to the effects of 
nonperturbative gluodynamics.

\section{Chromoelectric Polarizability}

Peskin specifies two conditions which permit the application of the operator
product expansion. The first is that the meson should be small, $r_Q^{-1}\gg 
\Lambda_{QCD}$, which implies $m_Q \gg \Lambda_{QCD}/\alpha_s(r_Q^{-1})$. 
The second constraint arises because gluons coupling to the heavy meson must arrange
themselves into colour singlets. Thus the emission of a single gluon -- which raises
the energy of the meson to that of an octet (or hybrid meson) state -- must be followed
quickly by a subsequent emission. The correlation time between these events is $\Delta t \sim
1/(E_a - E_\phi)$ where $E_a$ is the energy of the intermediate hybrid state. Thus the
colour singlet criterion is $E_a-E_\phi \sim \epsilon_B \gg \Lambda_{QCD}$ which imposes
the stronger constraint:

\begin{equation}
m_Q \gg n^2 \Lambda_{QCD}/\alpha_s^2.
\label{constraint}
\end{equation}

\noindent
We have introduced the Coulombic binding energy 
$\epsilon_B = m_Q C_F^2 \alpha_s^2/4$ ($C_F = (N_c^2-1)/(2 N_C)$) and the 
principle quantum number of the heavy meson, $n$. 
Eq. \ref{constraint} implies that the potential felt by the heavy quarks is
perturbative and hence that the heavy meson wavefunction is nearly Coulombic.
Peskin estimates that the condition of Eq. \ref{constraint} is met for $m_Q \gg 25$ GeV 
for $n=1$.
We note that this 
result is obtained in the large $N_c$ limit where $E_a$ tends to $2 m_Q$  and hence 
$\Delta t \sim 1/\epsilon_B$. An updated limit in which this constraint is 
considerably relaxed will be established in section IV.

Under the conditions specified above, gluon emissions must arrange themselves into small
colour singlet clusters which are attached to a small region in spacetime in which the heavy
meson is in an octet state. This observation permits the application of the operator
product expansion. Peskin applies this idea by exponentiating all possible two-gluon
couplings to the heavy meson. The result is a gauge invariant effective interaction of
the form

\begin{equation}
L_{eff} = -\sum_{N=1} c_E^{(N)\, ij} a_0^3 \epsilon_B^{2-2N}\cdot E^i D_0^{2N-2} E^j
\label{leff}
\end{equation}

\noindent
where $D_0$ is the temporal component of the covariant derivative.
We follow Peskin and introduce the dimensionful
parameters $a_0 = 2/(C_F \alpha_s m_Q)$ (the Coulombic Bohr radius) and $\epsilon_B$ to
make the Wilson coefficients, $c_E^{(N)}$, dimensionless. We note that the leading term 
has already been given in covariant form in Eq.~\ref{L0} and as a model in Eq.~\ref{leut}.

The expression for the Wilson coefficient is

\begin{equation}
c_E^{(N)\, ij} = {2 \pi \alpha_s \epsilon_B^{2N-2}\over N_c a_0^3} \langle \phi\vert r^i {1\over (H_a - E_\phi)^{2N-1}} r^j \vert \phi\rangle
\label{ceFull}
\end{equation}
and $\phi$ represents the heavy meson of interest. For S-wave states
$c_E^{(N)\, ij} = \delta^{ij} c_E^{(N)}$.  Finally, using 1s Coulombic wavefunctions and 
neglecting the adjoint potential yields the result\cite{Peskin} for $c_E^{(1)}(1s)$ given in 
Eq. \ref{ce} (we suppress the superscript from now on). A similar computation gives

\begin{equation}
c_E(2s) = {502\over 7} c_E(1s).
\label{ce2s}
\end{equation}

\noindent
In general $r_Q \sim n^2/(m_Q \alpha_s)$, and the energy denominator scales as 
$m_Q \alpha_s^2/n^2$, thus
$c_E(ns) \sim n^6$ (factors of $m_Q$ and $\alpha_s$ cancel against the 
prefactors in Eq.~\ref{ceFull}). It is clear that the OPE breaks down very quickly 
with the principle quantum number.

As we have already remarked, these estimates have been used to compute 
the strength of hadronic interactions in a variety of applications. However, 
a number of strong assumptions have been made in deriving them. Certainly, it is
not clear that the adjoint potential need be as simple as perturbation theory indicates.
Fortunately recent improvements in lattice gauge theory have allowed for an accurate
determination of this interaction in the heavy quark regime\cite{JKM}. It is therefore
expedient to confront the assumptions of Refs.\cite{Peskin,LMS,BM,HL,MV,KS} 
with lattice gauge theory in an attempt to establish the validity of Eqs.~\ref{ce} and 
\ref{leff}.
Thus we briefly review the current knowledge of hybrid potentials before moving
on to a re-evaluation of the polarizability and the operator product expansion itself.

\section{Adiabatic Hybrid Spectrum}

A simple consequence of the fact that glue is confined is that it must manifest itself
as a discrete spectrum in the presence of a static colour source and sink. In this case the 
physical hadrons are heavy hybrid mesons. It is 
relatively easy to study heavy hybrids by constructing gluonic configurations on the
lattice which are analogous to those of a diatomic molecule. Indeed, the gluonic 
configurations may be described with the same set of quantum numbers as diatomic
molecules: $\Lambda_\eta^Y$. Here the
projection of the total gluonic angular momentum onto the $Q\bar Q$ axis is denoted by
$\Lambda$ which may take on values $\Sigma$, $\Pi$, $\Delta$ = 0,1,2, etc. The combined
operation of 
charge and parity conjugation on the gluonic degrees of freedom is denoted by 
$\eta = u, g$
and $Y = \pm$ represents reflection of the system in a plane containing
the $Q\bar Q$ axis.  As with the diatomic molecule, all systems with 
$\Lambda$ greater than zero are doubly degenerate in $Y$.  Gluonic adiabatic
surfaces may be traced by allowing the heavy quark source and sink separation
to vary and hybrid mesons containing excited gluonic configurations may
be studied in the adiabatic Born-Oppenheimer approximation.

\begin{figure}[h]
\includegraphics[scale=0.35,angle=-90]{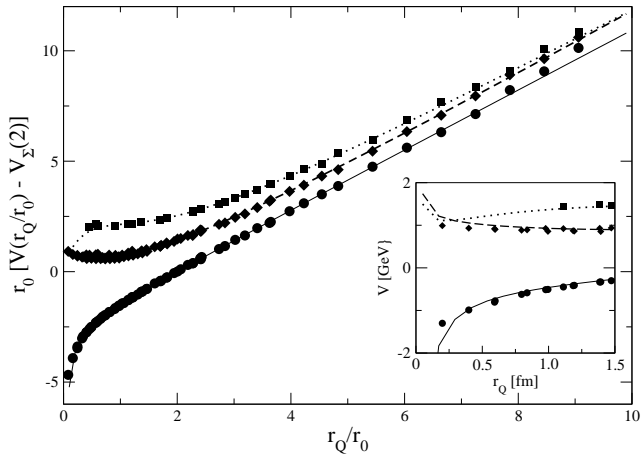}
\caption{\label{jkm}Low Lying Adiabatic Hybrid Surfaces. Lattice data for the
$\Sigma_g^+$ (circles), $\Pi_u$ (diamonds), and $\Pi_g$ (squares) surfaces from 
Ref.~\protect\cite{JKM}. Lines are simple parameterizations of the data.
The scale is $r_0 \approx 1/2$ fm. Inset: Potentials at Short Distance. The same
data is shown in traditional units. The $\Pi_u$ and $\Pi_g$ lines are those of the main figure 
with the addition of the perturbative adjoint potential $V_a$.}
\end{figure}

The results of a recent
lattice computation are presented in Figure \ref{jkm}\cite{JKM}. The lowest state is the $\Sigma_g^+$
surface and corresponds to the Wilson loop static interquark potential. The first (second)
excited state is the $\Pi_u$ ($\Pi_g$) surface and may be visualized as a gluonic flux tube with the
addition of a single `phonon'. A similar analogy exists for all of the higher states.

The inset
of the figure shows the lowest hybrid surfaces at scales less than 1.5 femtometres. 
The dashed and dotted lines are those of the main figure with the addition of the perturbative
adjoint potential, $V_a = \alpha_s/(6 r_Q)$\footnote{There is a subtlety here: the 
fitted value of the strong coupling is $C_F \alpha_s = 0.28$.
This is very close to the value expected for excitations of string-like degrees
of freedom, $\pi/12$, hence one suspects that the value of the strong coupling used
here does not represent the perturbative behaviour of QCD, but rather is an intermediate 
distance effect. Nevertheless, using this value for $\alpha_s$ overestimates the region 
of validity
of the operator product expansion, and the conclusions presented below stand.}. The figure
demonstrates that perturbative behaviour has not been seen at $r_Q \approx 0.2$ fm
(points)
and that this is consistent with expectations (dotted and dashed lines).

\section{Discussion}

We first note that the nonperturbative gluodynamics shown in Figure~\ref{jkm} indicates
that the octet-singlet splitting used in Eq.~\ref{constraint} is not accurately described by
the Coulombic binding energy. Rather, the figure indicates that this splitting is 
roughly 1 GeV at typical hadronic scales. Thus 
the constraint is not nearly as strong at finite $N_c$ and with reasonable hybrid 
potentials. One concludes that gluons are largely correlated in time as 
required, lending hope to the idea that the application of the OPE to hadronic 
interactions may be unexpectedly robust. Unfortunately, a new constraint exists, which we
now demonstrate.

A central criterion for the validity of the operator product expansion is that the hybrid
surfaces shown in Figure~\ref{jkm} approach a universal form at short distances. 
Figure~\ref{jkm} makes it clear that a
universal hybrid potential behaviour (namely $V_a$) does not appear until

$$V_a(r_Q) \gg V_{{\Lambda_\eta^Y}'}(r_Q) - V_{\Lambda_\eta^Y}(r_Q).$$

\noindent
Since the typical hybrid surface separation is order $\Lambda_{QCD}$ for small $r_Q$ (and is much
larger for the splitting relevant to ground state mesons), one has 
$m_Q \gg 6 \Lambda_{QCD}/\alpha_s^2 \approx 150$ GeV.
Thus Eq. \ref{constraint} is recovered (albeit with an unlucky additional large factor);
however, the constraint is now a {\it necessary condition} rather than merely sufficient as
before. Thus, although the condition which insures the emission of colour singlet states is 
likely to be satisfied for all quark masses, the hidden assumption in the method, namely
that a universal octet potential is relevant, is only true for very heavy quarks.

This conclusion has a simple interpretation in Leutwyler's random field model: the 
appearance of a discrete hybrid spectrum makes it clear that the correct representation
(here we employ the Born-Oppenheimer approximation)
of the matrix element of $\delta H$ of Eq.~\ref{leut} is as follows

\begin{eqnarray}
\delta E_n &=& \langle \phi_n | \delta H | \phi_n \rangle \nonumber \\
 &\to& \langle \phi_n; \Sigma_g^+ | \delta H | \phi_n; \Sigma_g^+\rangle \nonumber \\
&=& \sum_{h, \Lambda, \eta, Y} {\vert\langle \phi_n; \Sigma_g^+ | {\bf E}\cdot {\bf r}|h; \Lambda_\eta^Y\rangle \vert^2 \over (E_h(\Lambda_\eta^Y) - E_\phi)}.
\end{eqnarray}

\noindent
In this expression $h$ represents all of the nongluonic quantum numbers which describe an
intermediate heavy hybrid state. The essence of the operator product expansion is that this
expression {\it factorizes} ({\it ie.}, the Wilson coefficients depend on short range physics
only). Factorization requires that the hybrid energies in the denominator do not depend on
the gluonic quantum numbers, $\Lambda_\eta^Y$. It is only in this circumstance that 
the expression simplifies:

\begin{equation}
\delta E_n = \sum_h {\vert \langle \phi_n | {\bf r} | h \rangle \vert^2 \over (E_h - E_\phi)}
\cdot \langle \Sigma_g^+ \vert {\bf E}^2 \vert \Sigma_g^+ \rangle
\end{equation}

\noindent
and the operator product formalism is valid.

%
%
%
%
%


Finally, we consider the value of the chromoelectric polarizability in light of the lattice
hybrid data of Figure~\ref{jkm}. We choose to numerically compute the $\phi$ wavefunction
in the Born-Oppenheimer approximation with the aid of the 
lattice $\Sigma_g^+$ surface. The sum over intermediate hybrid states is performed
numerically by expanding in the eigenstates of the $\Pi_g$ surface (this is the lowest
surface which couples to a vector heavy hybrid -- which is the case we consider in the 
following).

\begin{figure}[h]
\includegraphics[scale=0.3, angle=-90]{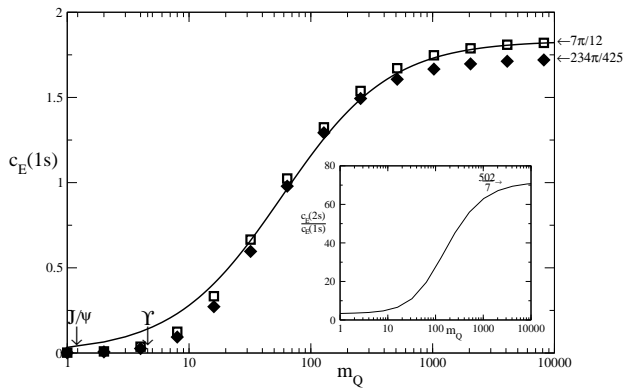}
\caption{\label{cePlot}Chromoelectric Polarizability as a Function of Quark Mass in GeV. 
Points represent $c_E(1s)$ as computed with the $\Pi_g$ surface with (diamonds)
and without (open squares) the perturbative adjoint potential, $V_a$. The line
is the approximation of Eq. \ref{Capprox}. Inset: The Ratio $c_E(2s)/c_E(1s)$ versus
Quark Mass.}
\end{figure}

The results are shown as the open squares in Figure~\ref{cePlot}.  One sees
that the Peskin result of Eq.~\ref{ce} is recovered (arrow) in the very heavy quark mass
limit, as expected\footnote{We note that the approach to the Coulombic limit is, in part,
very slow because of the small value of the strong coupling. For example, it is much
faster if typical quark model values for the strong coupling are employed.}. However, the 
value of $c_E(1s)$ at the $\Upsilon$ or $J/\Psi$ masses (arrows on the abscissa) is
highly suppressed with respect to the asymptotic value. The diamonds are numerically
obtained values for the polarizability in the case that the adjoint potential has been
included in the $\Pi_g$ surface parameterization. Again, the analytical result, $c_E(1s) = 
234 \pi/425$\cite{HL}, is approached very slowly in quark mass.

It is tempting to speculate that the majority of the finite quark mass correction to 
Peskin's result is due to the hybrid mass gap. Allowing for this in Eq.~\ref{ceFull}
yields the following generalization of Eq.~\ref{ce}:

\begin{eqnarray}
c_E(1s) &=& {8 \pi \over 3 (N_c^2-1) v^6} \Big( 256(1+v)^{3/2} - 256\nonumber \\
&-& 384 v - 96 v^2 + 16 v^3 - 6 v^4 + 3 v^5 \Big) 
\label{Capprox}
\end{eqnarray}
where $v = {4 V_0 \over (C_F \alpha_s)^2 m_Q}$ and $V_0$ is the strength of the relevant
hybrid potential at
its minimum. The resulting expression is shown as a solid line in Figure~\ref{cePlot};
evidently the agreement is quite good and this expression may serve as a useful
extrapolation to light quark masses.

Finally we display the ratio $c_E(2s)/c_E(1s)$ in the inset of Fig.~\ref{cePlot}. Again
the Peskin result (Eq.~\ref{ce2s}) is approached only very slowly in the heavy quark limit.
We note that Eq.~\ref{ce2s} leads to the uncomfortable prediction that 
the $\Upsilon'$ interacts 5000
times more strongly with nuclear matter than does the $\Upsilon$. The inset shows, however, 
that this prediction is substantially moderated (from 5000 to roughly 15) when finite
quark mass effects are taken into account.

\section{Conclusions}

According to the arguments of Refs \cite{Peskin,LMS,HL,MV}, the electric polarizability of 
a small meson
controls the strength of its interactions with hadronic matter via the operator product
expansion. We have recomputed this strength with the aid of lattice hybrid potentials\footnote{
The lattice hybrid potentials employed here were obtained in the Born-Oppenheimer
approximation, wherein gluonic degrees of freedom respond rapidly to slow quark motion.  
It may be shown that the requirements for the validity of the Born-Oppenheimer approximation
coincide with Eq. \ref{constraint}.
}
and find that the 
large mass gap between the ground state ($\Sigma_g^+$) and excited
state gluonic configurations leads to a strong suppression of the electric polarizability
as the quark mass is reduced.  The result is that, if one neglects issues of the applicability of
the OPE, previous estimates of interaction strengths are reduced by roughly a factor of 100.
Thus the arguments of Khazeev and Satz concerning the utility of  $J/\psi$ suppression as
a quark-gluon plasma diagnostic are strengthened. Alternatively non-quark-exchange  
$J/\psi$--nuclear matter interactions are greatly reduced, suggesting that quark 
exchange mechanisms should be
carefully considered in the analysis of the $J/\psi$--nuclear matter binding issue.

On a more general level
we have argued that the assumptions underlying the operator product expansion description
of heavy hadron interactions are violated for all physical states. This situation arises
because the sufficient condition on temporal correlations among gluons has been 
replaced with a necessary condition on the applicability of factorization which is only
true for very heavy quarks.

The short length scale required for factorization arises for a number of reasons.
Certainly the fact that the strong coupling is not large and that the ratio of fundamental
and adjoint Casimirs is also small, help to undermine the reliability of the OPE. However,
the relatively flat behaviour of the adiabatic hybrid surfaces below 1 fm must be considered
the leading cause. One may speculate that this arises due to the robust persistence of 
string-like field configurations, even at quite small interquark separations.  It thus appears
that strong nonperturbative gluodynamics conspires to bring about the demise of the operator
product formalism in this application. 

Although we have said nothing about the utility of
the OPE in the very heavy quark limit, the authors of Ref. \cite{FK} show that the interaction
between very small colour dipoles becomes nonperturbative (it is essentially correlated two
pion exchange). It thus appears that the premise of the OPE and any effective field theoretic
approach to the interactions of small hadrons is compromised.

\begin{acknowledgments}

We thank C.~J. Morningstar for extensive discussions, K. Waidelich for assistance during early
stages of this research, and N. Brambilla and D. Kharzeev for useful comments.
This work was supported by the US Department of Energy under contracts
DE-FG02-00ER41135  and  DE-AC05-84ER40150 and by a Warga Fellowship (OL).
\end{acknowledgments}


\end{document}